\documentclass[12pt]{article}
\usepackage[utf8]{inputenc}
\usepackage[export]{adjustbox}
\usepackage{authblk}
\usepackage{bbm}
\usepackage{pgfpages}
\usepackage{graphicx}
\usepackage{multicol}
\usepackage{csquotes}
\usepackage{multirow}
\usepackage{graphicx} 
\usepackage{booktabs} 
\usepackage{authblk}
\usepackage{caption}
\usepackage{subcaption}
\usepackage{stackengine}
\usepackage{amsmath}

\usepackage[margin=1in]{geometry}
\usepackage{setspace}

\usepackage[maxfloats=256]{morefloats}

\usepackage{comment}
\usepackage{lscape}
\usepackage{hyperref}
\hypersetup{
    colorlinks=true,
    linkcolor=blue,
    filecolor=blue,      
    urlcolor=blue,
    citecolor=blue,
}
\usepackage[capposition=top]{floatrow}

\usepackage{amssymb}
\usepackage{stackrel}
\usepackage{natbib} % bibliography

\title{In-between Transatlantic \textit{(Monetary)} Disturbances}

\author[1]{Jeanne Aublin}

\affil[1]{McGill University}

\author[2]{Santiago Camara}

\affil[2]{McGill University \& Red-NIE}
\date{\normalsize This Draft \today}

\begin{document}
    
\maketitle

\begin{abstract}
    This paper studies the spillovers of European Central Bank (ECB) interest rate shocks into the Canadian economy and compares them with those of the U.S. Federal Reserve (Fed). We combine a VAR model and local projection regressions with identification strategies that explicitly purge information effects around policy announcements. We find that an ECB rate hike leads to a depreciation of the Canadian dollar and a sharp contraction in economic activity. The main transmission channel is international trade: ECB shocks trigger a decline in oil prices and exports, while leaving domestic financial conditions largely unaffected. By contrast, Fed shocks tighten Canadian financial conditions significantly, with more limited effects on trade flows. These findings show that Canada is exposed to foreign monetary policy both directly and indirectly, through its integration in global financial and trade markets.

    \medskip

    \medskip

    \medskip

    \medskip
    
\end{abstract}

%%%%%%%%%%%%%%%%%%%%%%%%%%%%%%%%%%%%%%%%%%%%%%%%%%%%%%%%%%%%%%%%%%%%%%
%% Introduction
\newpage
\section{Introduction} \label{sec:introduction}

Canada has long served as the textbook example of a small, open economy (see \cite{mendoza1991real, coletti2008inflation, justiniano2010monetary, uribe2017open}). Its limited influence on global financial conditions and reliance on international trade make it an ideal setting to study the international transmission of monetary policy. Most empirical work, however, summarizes Canada’s external environment using U.S.\ variables alone, reflecting the deep trade and financial ties between the two countries (see \cite{devereux2003monetary, barakchian2015transmission}). This U.S.-centric view overlooks an important fact: interest rate shocks from the European Central Bank (ECB) also generate sizable spillovers into the Canadian economy. These spillovers operate mainly through trade and commodity price channels, and are strong enough to trigger macroeconomic fluctuations in Canada comparable to those caused by U.S.\ shocks.

Understanding such spillovers is of first-order importance for both policymakers and researchers. Central banks in small open economies like Canada operate in an environment increasingly shaped by the actions of major central banks abroad. While the influence of the U.S.\ Federal Reserve (Fed) on Canada has been extensively documented, the role of ECB policy has received relatively little attention. This omission is noteworthy given Canada’s dependence on commodity exports and the global reach of European demand conditions, which can affect Canadian activity through trade networks, commodity markets, and shifts in financial sentiment.

\begin{figure}[ht]
  \centering
  \begin{subfigure}[b]{0.48\textwidth}
    \includegraphics[width=\textwidth]{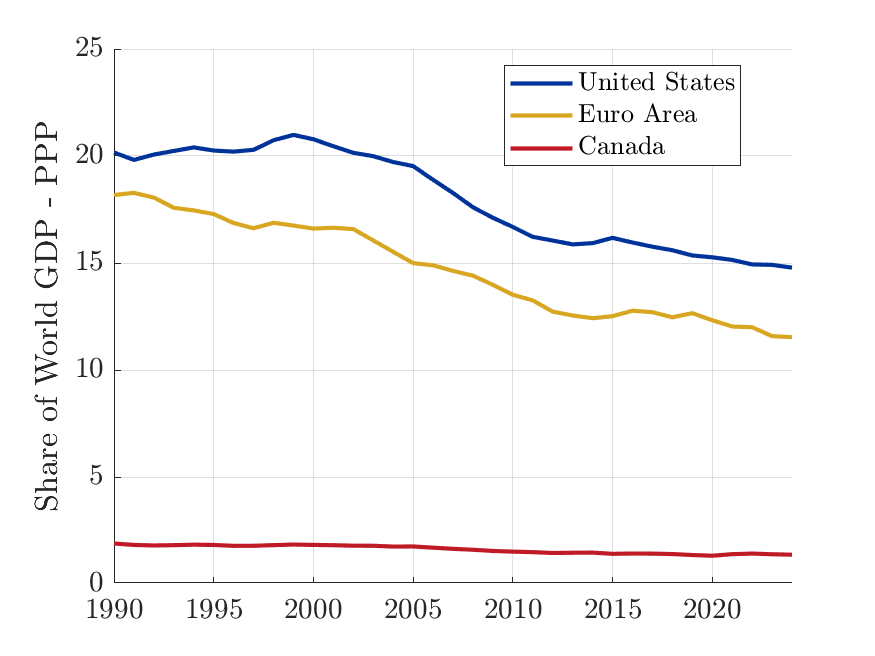}
    \caption{Share of World GDP (PPP)}
  \end{subfigure}
  \hfill
  \begin{subfigure}[b]{0.48\textwidth}
    \includegraphics[width=\textwidth]{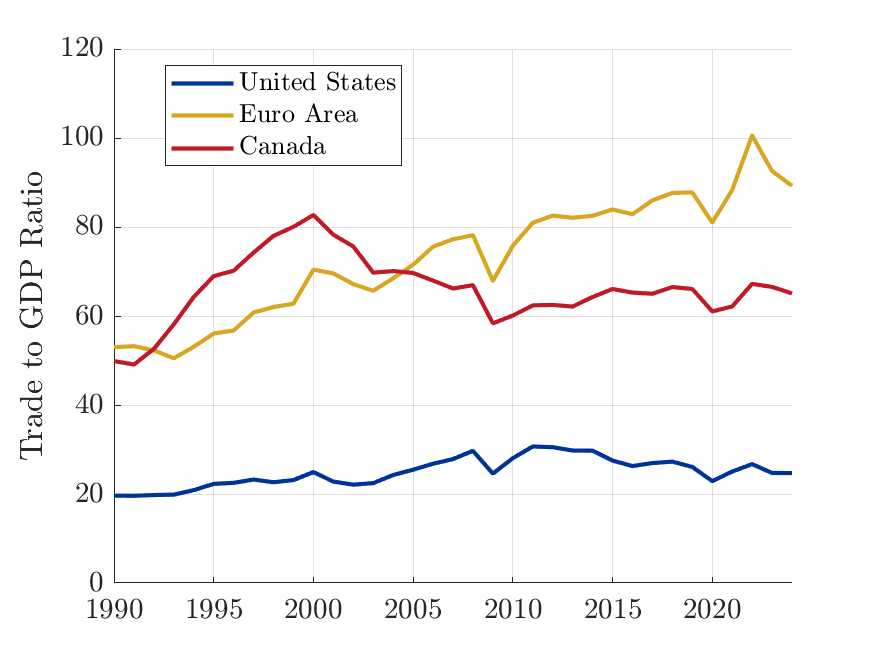}
    \caption{Trade-to-GDP Ratios}
  \end{subfigure}
  
  \vspace{0.5em}
  \begin{subfigure}[b]{0.48\textwidth}
    \includegraphics[width=\textwidth]{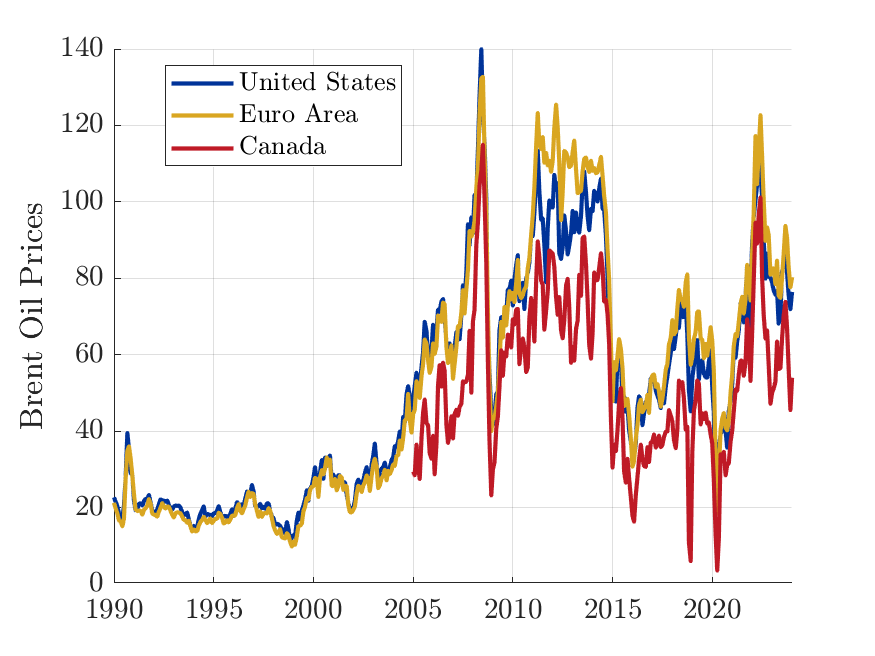}
    \caption{Oil Price Correlation (WCS, Brent, WTI)}
  \end{subfigure}
  \hfill
  \begin{subfigure}[b]{0.48\textwidth}
    \includegraphics[width=\textwidth]{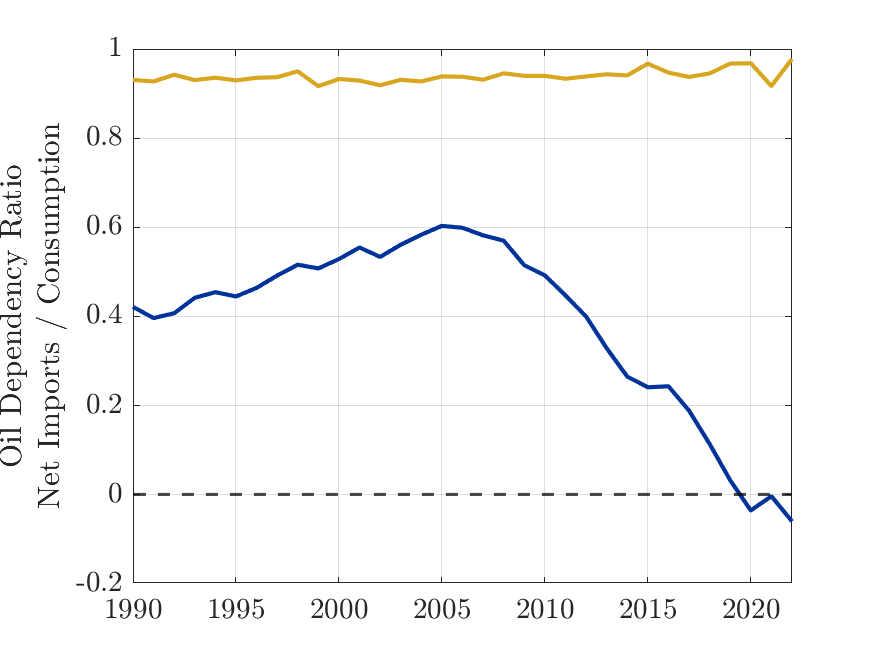}
    \caption{Oil Dependency Ratio (Net Imports / Consumption)}
  \end{subfigure}

  \caption{Global Relevance and Exposure: Canada, the U.S., and the Euro Area}
  \label{fig:Intro_Global_Exposure}
\end{figure}

At first glance, the Euro Area might appear peripheral to Canada. Bilateral trade flows are modest—less than 8\% of Canadian exports are destined for the Euro Area, and imports from the bloc account for a similarly small share. Financial linkages are also limited compared to those with the United States or the Americas. Yet the Euro Area is the world’s second-largest economic bloc, accounting for nearly 15\% of global GDP, and is far more open to trade than the United States. As shown in Figure~\ref{fig:Intro_Global_Exposure}, Euro Area trade has consistently exceeded 80\% of GDP, compared to 25\% in the U.S. and 60--70\% in Canada. This greater openness magnifies the global reach of Euro Area demand fluctuations, particularly in internationally traded sectors such as commodities.

One of the clearest channels through which Canada can be exposed to Euro Area developments is oil prices. Oil is Canada’s largest export and a key driver of its terms of trade. Panel~(c) of Figure~\ref{fig:Intro_Global_Exposure} shows that the Canadian benchmark (Western Canadian Select) tracks global benchmarks such as Brent (the European standard) and West Texas Intermediate (the U.S. benchmark) closely. This implies that Canadian export revenues depend on global oil market dynamics, regardless of the ultimate destination of shipments. Importantly, while the U.S.\ has recently become a net exporter of oil, the Euro Area remains a large net importer (Panel~(d)). As a result, oil prices—and hence Canadian export revenues—are more sensitive to demand shocks originating in the Euro Area than in the U.S. A tightening of ECB policy that reduces European energy demand may lower global oil prices, indirectly affecting Canadian GDP, the exchange rate, and fiscal revenues.

More broadly, recent research highlights the global reach of non-U.S.\ central banks. ECB monetary policy has been shown to affect global financial conditions, sovereign spreads, and real activity outside Europe, while the People’s Bank of China (PBoC) increasingly influences commodity markets, capital flows, and macroeconomic outcomes in emerging economies (see \cite{miranda2020global, miranda2022global}). These findings underscore the need to move beyond a U.S.-centric view of global monetary spillovers.

\medskip
\noindent
\textbf{Contributions.} This paper provides systematic evidence that ECB monetary policy shocks generate sizable macroeconomic fluctuations in Canada. Using high-frequency monetary policy surprises combined with VAR and local projection regressions, we show that ECB rate hikes lead to a sharp depreciation of the Canadian dollar, lower GDP, and reduced exports. The effects operate primarily through commodity and trade channels, contrasting with the financial channel that dominates Fed spillovers. Notably, the magnitude of ECB spillovers is comparable to those triggered by U.S.\ monetary policy shocks. Our findings challenge the common practice of modeling Canada’s external environment solely through U.S.\ variables, and call for a broader perspective that recognizes multiple sources of foreign monetary influence.

\bigskip
\noindent
\textbf{Organization.} Section \ref{sec:data_model} describes the dataset, identification strategy, and econometric specification. Section \ref{sec:transatlantic_spillovers} presents the benchmark results. Section \ref{sec:transatlantic_spillovers_transmission_channels} compares the transmission channels of ECB and Fed spillovers. Section \ref{sec:additional_robustness_checks} reports robustness checks. Section \ref{sec:conclusion} concludes.

\bigskip
\noindent
\textbf{Related literature.} Our work connects to three strands of research. First, an extensive literature studies how U.S.\ monetary policy shocks affect Canada, reflecting the two countries’ deep integration. Early contributions include \citet{cushman1997identifying}, who develop a small open economy VAR showing that U.S.\ monetary policy has significant effects on Canadian output and the exchange rate. More recent work analyzes the impact on Canadian asset prices \citep{li2010impact} and regional heterogeneity across provinces \citep{potts2010variations}. We contribute by revisiting Fed spillovers to Canada using high-frequency identification.

Second, we build on research emphasizing the dual nature of monetary announcements, which contain both policy and information components. Neglecting this distinction biases estimates of real effects. Recent work, including \citet{camara2025spillovers}, shows that separating monetary policy shocks from information shocks yields markedly different spillover patterns, particularly for emerging markets. Related studies include \citet{degasperi2020global} and \citet{camara2024international}. We contribute by showing that, even in Canada, controlling for information effects is essential for a consistent assessment of spillovers.

Third, we relate to the literature on the global impact of ECB monetary policy. \citet{jarocinski2022central} document “transatlantic” spillovers of Fed and ECB policy to the U.S.\ and Euro Area, while \citet{miranda2022tale} highlights common propagation channels through trade and global risk-taking. We extend this line by showing that the Canadian economy, despite limited direct Euro Area exposure, is significantly affected by ECB rate shocks—primarily via oil prices and trade flows, in contrast to the financial-channel transmission of Fed shocks.\footnote{An alternative literature studies Euro Area linkages more broadly, e.g., \cite{dees2007exploring}, but does not focus on monetary policy shocks.}

%%%%%%%%%%%%%%%%%%%%%%%%%%%%%%%%%%%%%%%%%%%%%%%%%%%%%%%%%%%%%%%%%%%%%%
\section{Data Description \& Econometric Specification} \label{sec:data_model}

This section describes the dataset, the construction of monetary policy shocks, and the econometric framework used to estimate impulse responses. Section \ref{subsec:data_and_event} outlines the Canadian and international variables and the identification strategy for policy shocks. Section \ref{subsec:econometric_specification} then presents the local projection and VAR specifications.

%%%%%%%%%%%%%%%%%%%%%%  %%%%%%%%%%%%%%%%%%%%%%%%%%%%%%%%%%%%%%%%%%%%%%%%
\subsection{Data Description \& Identification Strategy} \label{subsec:data_and_event}

\noindent
\textbf{Dataset.} 
Most Canadian variables are sourced from the ``Large Canadian Database for Macroeconomic Analysis'' compiled by \citet{fortin2022large}, with the exception of the CAD/EUR exchange rate, which is obtained from the IMF-IFS database. U.S.\ variables are primarily drawn from the St.\ Louis FRED and the Federal Reserve,\footnote{The only variable taken directly from the Federal Reserve website is the updated Excess Bond Premium of \citet{gilchrist2012credit}, available at \url{https://www.federalreserve.gov/econres/notes/feds-notes/updating-the-recession-risk-and-the-excess-bond-premium-20161006.html}.} while Euro Area variables are obtained from FRED and Eurostat.\footnote{Eurostat series retrieved from \url{https://ec.europa.eu/eurostat/web/main/data/database}, specifically indicator ``sts\_inpr\_m'' under \textit{Production in industry / Short-term business statistics}.}

The benchmark sample includes five Canadian variables and four foreign variables that together capture economic activity, prices, monetary policy, and financial conditions. For Canada, we use the nominal exchange rate (against the U.S.\ dollar when analyzing Fed shocks and against the euro when analyzing ECB shocks), the Bank of Canada’s policy rate, the Consumer Price Index (CPI), an equity price index, and the monthly GDP index.\footnote{We use GDP for Canada, but industrial production for the U.S.\ and Euro Area due to data availability. Since manufacturing represents a smaller share of Canadian GDP, we consider monthly GDP a more comprehensive measure of Canadian activity. In Section \ref{sec:additional_robustness_checks}, we show that results are robust to using Canada’s industrial production index instead.} For the United States, the dataset includes the federal funds rate, the industrial production index, the PCE price index, and the Excess Bond Premium (EBP) of \citet{gilchrist2012credit}. For the Euro Area, we use the ECB’s main refinancing rate, the industrial production index, the harmonized consumer price index (HICP), and the ICE BofA Euro High Yield Index Option-Adjusted Spread (BAMLHE00EHYIOAS). This set of variables closely follows the international spillover literature, such as \citet{camara2025spillovers}.

\begin{figure}[ht]
     \centering
     \begin{subfigure}[b]{0.485\textwidth}
         \centering
         \includegraphics[width=\textwidth]{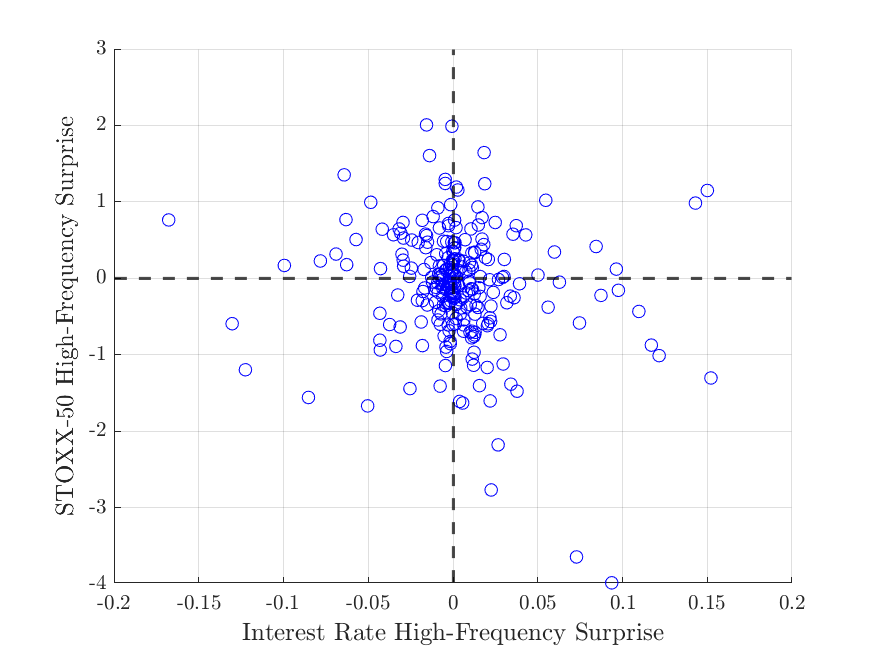}
         \caption{ECB Meetings}
         \label{fig:ECB_Surprises}
     \end{subfigure}
     \begin{subfigure}[b]{0.485\textwidth}
         \centering
         \includegraphics[width=\textwidth]{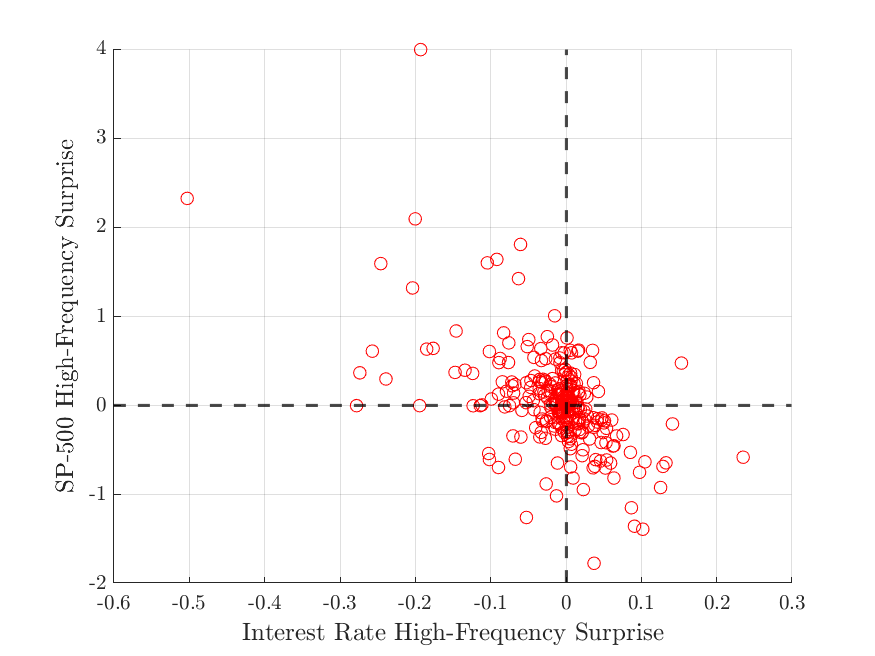}
         \caption{Fed Meetings}
         \label{fig:Fed_Surprises}
     \end{subfigure}
    \caption{High-frequency surprises in interest rates and stock prices around policy announcements}
    \label{fig:Surprises}
\end{figure}

\medskip
\noindent
\textbf{Identification of monetary policy shocks.} 
Our benchmark identification strategy follows the high-frequency approach of \citet{jarocinski2020deconstructing}, which separates monetary policy shocks from central bank information effects by examining the joint behavior of asset price surprises around policy announcements. The logic is straightforward: a contractionary monetary policy shock should raise short-term interest rates while lowering equity prices, producing a negative co-movement. By contrast, when interest rates and equity prices move in the same direction, the announcement is more likely to convey information about the economic outlook rather than reflect a pure policy innovation. Figure \ref{fig:Surprises} shows the underlying high-frequency surprises in interest rates and equity prices around ECB (blue) and Fed (red) announcements. From these figures, it is clear that multiple announcements are associated with a co-movement at odds with conventional theory.

To recover the underlying structural shock, the identification strategy uses the rotational–angle decomposition proposed by \citet{jarocinski2022central}, which separates monetary policy shocks from central bank information shocks using high-frequency asset price surprises around policy announcements. The intuition is that movements in interest rate and equity futures around announcement windows can be explained by two underlying factors: a pure monetary policy shock and a central bank information disclosure shock. The rotational-angle method recovers these two orthogonal components by searching over possible rotations of the data that are consistent with the set of sign restrictions. 

\begin{figure}[ht]
    \centering
    \includegraphics[width=0.75\linewidth]{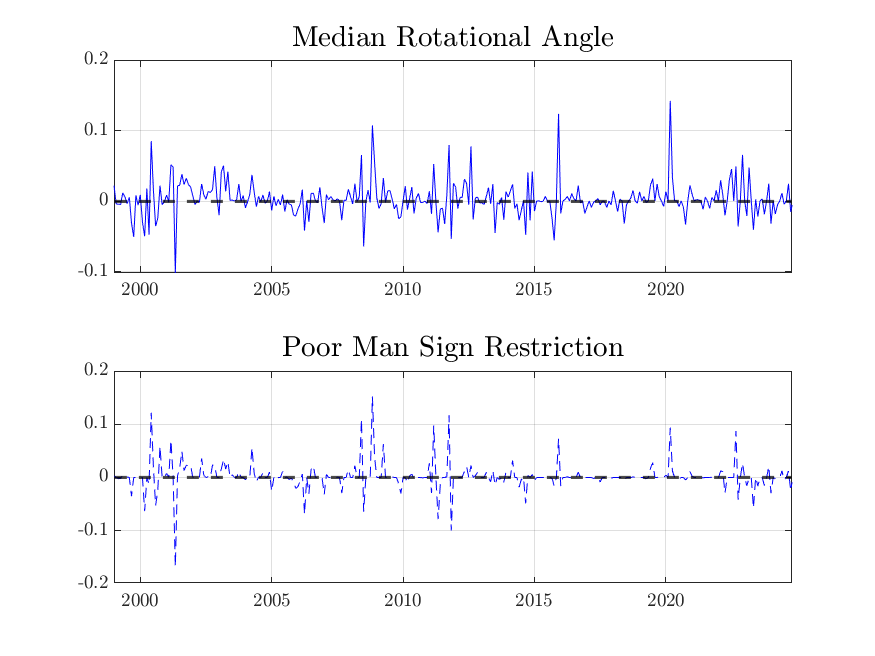}
    \caption{ECB Pure Monetary Policy Shocks (Median Rotation)}
    \label{fig:ECB_Shocks}
\end{figure}

\begin{figure}[ht]
    \centering
    \includegraphics[width=0.75\linewidth]{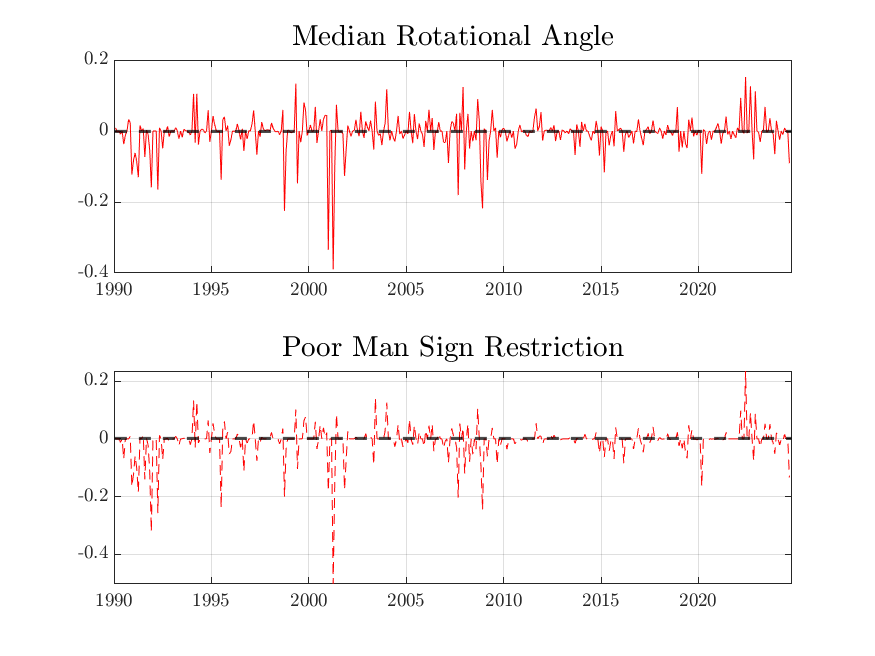}
    \caption{Fed Pure Monetary Policy Shocks (Median Rotation)}
    \label{fig:Fed_Shocks}
\end{figure}

Because this procedure delivers a range of admissible decompositions, we follow \citet{jarocinski2022central} in using the median rotation as our benchmark definition of a pure monetary policy shock. Figures \ref{fig:ECB_Shocks} and \ref{fig:Fed_Shocks} display the resulting shock series for the ECB and the Fed. As a robustness check, we also implement the ``Poor Man’s Sign Restriction'' (PMSR) of \citet{jarocinski2020central}. This simpler classification relies only on the \emph{sign} of the co-movement between interest rate and equity surprises on announcement days: a negative co-movement is interpreted as a monetary policy shock, while a positive co-movement is classified as an information disclosure shock. Section \ref{sec:additional_robustness_checks} reports results under this alternative approach, which confirm the robustness of our main findings.

%%%%%%%%%%%%%%%%%%%%%%%%%%%%%%%%%%%%%%%%%%%%%%%%%%%%%%%%%%%%%%%%%%%%%%
\subsection{Econometric Specification} \label{subsec:econometric_specification}

This subsection outlines the benchmark econometric specifications used to estimate the impulse responses to monetary policy shocks presented in Section \ref{sec:transatlantic_spillovers}. We rely on two complementary approaches: Bayesian VARs and local projections \citep{jorda2005estimation}. Both identify the same population response under standard assumptions \citep{plagborg2021local}, but differ in their small-sample properties: VARs typically yield lower variance but may be more biased, whereas local projections are more flexible but less efficient \citep{olea2025local}. Using both approaches allows us to strike a balance between efficiency and robustness.

\medskip
\noindent
\textbf{Bayesian VARs.}  
We estimate a Bayesian Vector Autoregression (BVAR) with three lags of endogenous variables and include the identified ECB and Fed monetary policy shocks as exogenous regressors. The specification also includes a linear time trend and a dummy for the COVID-19 period (March 2020–June 2021) to account for pandemic-related disruptions.\footnote{Lag length is chosen as a compromise between the Akaike Information Criterion (four lags) and the Bayesian Information Criterion (two lags). Robustness to alternative lag choices is reported in Section \ref{sec:additional_robustness_checks}.} 

We adopt a Normal-Wishart prior following \citet{dieppe2016bear}, with standard hyperparameter values: prior mean for the first own lag set to 0.8, overall tightness ($\lambda_1$) set to 0.1, and lag decay ($\lambda_3$) equal to 1. The exogenous monetary policy shocks are assigned diffuse priors.

\medskip
\noindent
\textbf{Local projections.}  
As a complementary approach, we estimate impulse responses using the local projection framework:
\begin{align}
    y_{t+h} &= \beta^{Shock}_{h} i^{\text{Shock}}_t 
    + \sum^{J_i}_{j=1} \phi^{Shock}_i i^{\text{Shock}}_{t-j} 
    + \sum^{J_y}_{j=1} \delta^{j}_i y_{t-j} 
    + \sum^{J_x}_{j=1} \alpha^{j}_i x_{t-j} 
    + \sum^{J^{*}_x}_{j=1} \alpha^{j,*}_i x^{*}_{t-j} \notag \\
    &\quad + \gamma_1 \cdot \text{Trend}_t 
    + \gamma_2 \cdot \text{CovidDummy}_t 
    + \epsilon_{t}, \label{eq:LP_Benchmark}
\end{align}
where $y_{t+h}$ denotes the outcome variable $h$ months after $t$, and $\beta^{Shock}_{h}$ traces the response to the contemporaneous monetary policy shock $i^{\text{Shock}}_t$. We include three lags of the dependent variable, Canadian control variables $x$, foreign variables $x^{*}$, and the shock series, selected based on the Akaike Information Criterion. Lagged dependent variables capture seasonal dynamics, while lagged shocks improve efficiency as emphasized by \citet{ramey2016macroeconomic} and \citet{jorda2023local}. A linear time trend and a COVID-19 dummy control for low-frequency movements and pandemic-specific distortions. This specification is standard in the international monetary policy literature, including \citet{ilzetzki2021puzzling} and \citet{camara2025spillovers}. Section \ref{sec:additional_robustness_checks} presents robustness to alternative lag structures and model variations.

%%%%%%%%%%%%%%%%%%%%%%%%%%%%%%%%%%%%%%%%%%%%%%%%%%%%%%%%%%%%%%%%%%%%%%
\section{Transatlantic Monetary Spillovers} \label{sec:transatlantic_spillovers}

\begin{figure}[p]
    \centering
    \includegraphics[height=15cm,width=15cm]{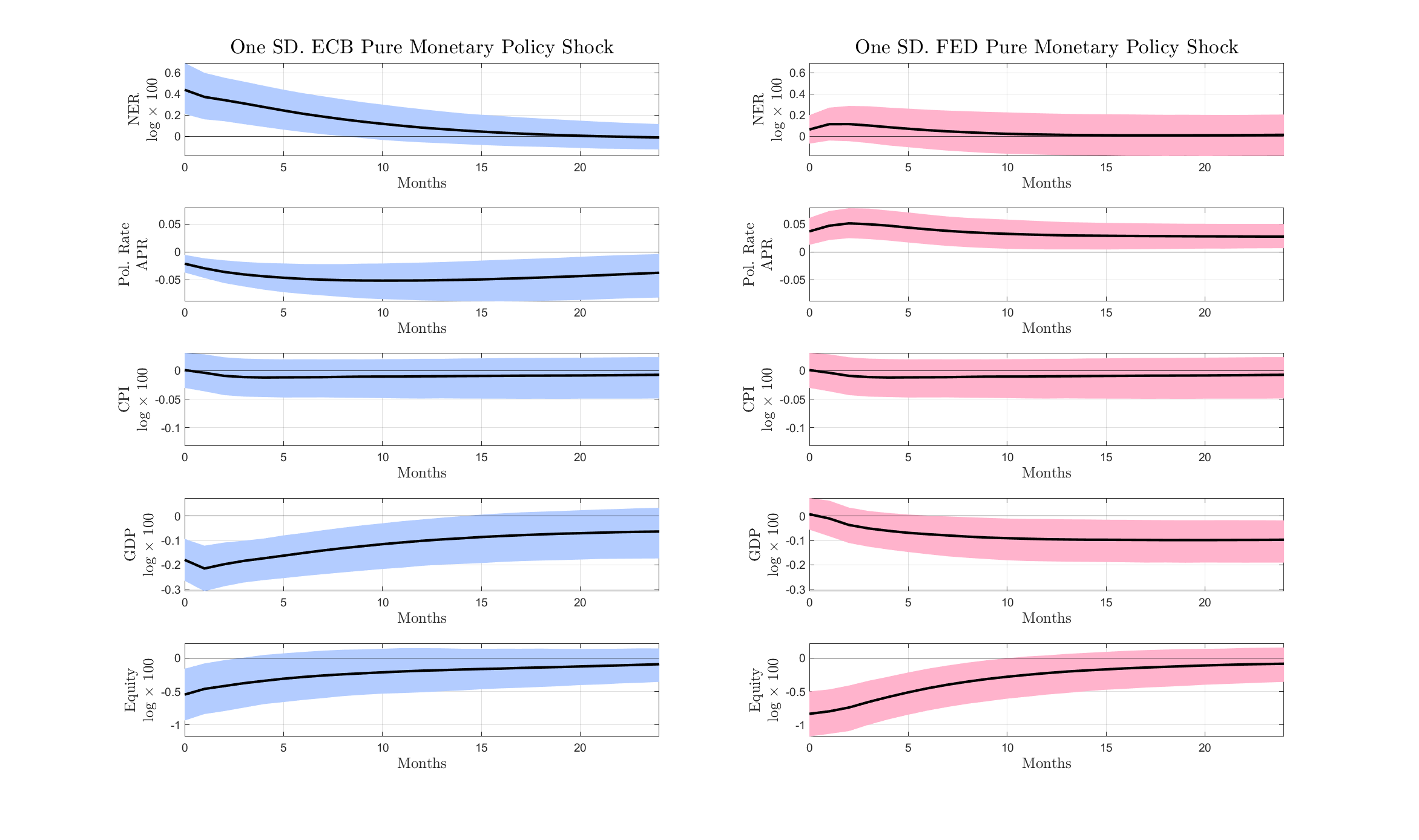}
    \caption{ECB \& Fed Interest Rate Shocks \\ Median Rotational Angle}
    \label{fig:Comparison_JK_MP_Median}
    \floatfoot{\textbf{Note:} Left panels correspond to ECB shocks; right panels to Fed shocks. The black line denotes the median IRF; shaded areas represent 90\% posterior credible intervals.}
\end{figure}

This section presents our main empirical findings. We estimate impulse response functions (IRFs) of Canadian macroeconomic variables to pure monetary policy shocks originating from the European Central Bank (ECB) and compare them to the effects of analogous shocks from the U.S. Federal Reserve (Fed). Figure \ref{fig:Comparison_JK_MP_Median} displays the IRFs identified via the median rotational angle: the left column (blue) reports responses to ECB shocks. In contrast, the right column (red) reports responses to Fed shocks.

A contractionary ECB shock generates an immediate and significant depreciation of the Canadian dollar against the euro. In response, the Bank of Canada lowers its policy rate persistently, cushioning the shock’s financial impact. Consumer prices decline only modestly and briefly, but real activity falls sharply: GDP contracts on impact and remains below trend for over a year. Equity prices also drop immediately, though this effect is short-lived and statistically indistinguishable from zero after four months.

By contrast, Fed shocks produce a markedly different adjustment. The Canadian dollar depreciates much less, as the Bank of Canada raises rather than lowers its policy rate. This monetary tightening is immediate and persistent, lasting up to 18 months. The GDP response is slower but longer-lasting: output begins to contract significantly only after six months, and the decline remains persistent for up to two years. Equity markets react much more strongly than under ECB shocks, falling sharply on impact and staying depressed for an extended period.

Taken together, these results highlight distinct transmission mechanisms. ECB shocks trigger an immediate and pronounced downturn in Canadian activity, operating primarily through trade and real channels, with limited effects on domestic financial conditions. Fed shocks, in contrast, tighten financial conditions directly, with persistent effects on equity markets and a more gradual but sustained contraction in output. This contrast motivates the more granular exploration of transmission channels in Section \ref{sec:transatlantic_spillovers_transmission_channels}.

\begin{figure}[p]
    \centering
    \includegraphics[height=15cm,width=15cm]{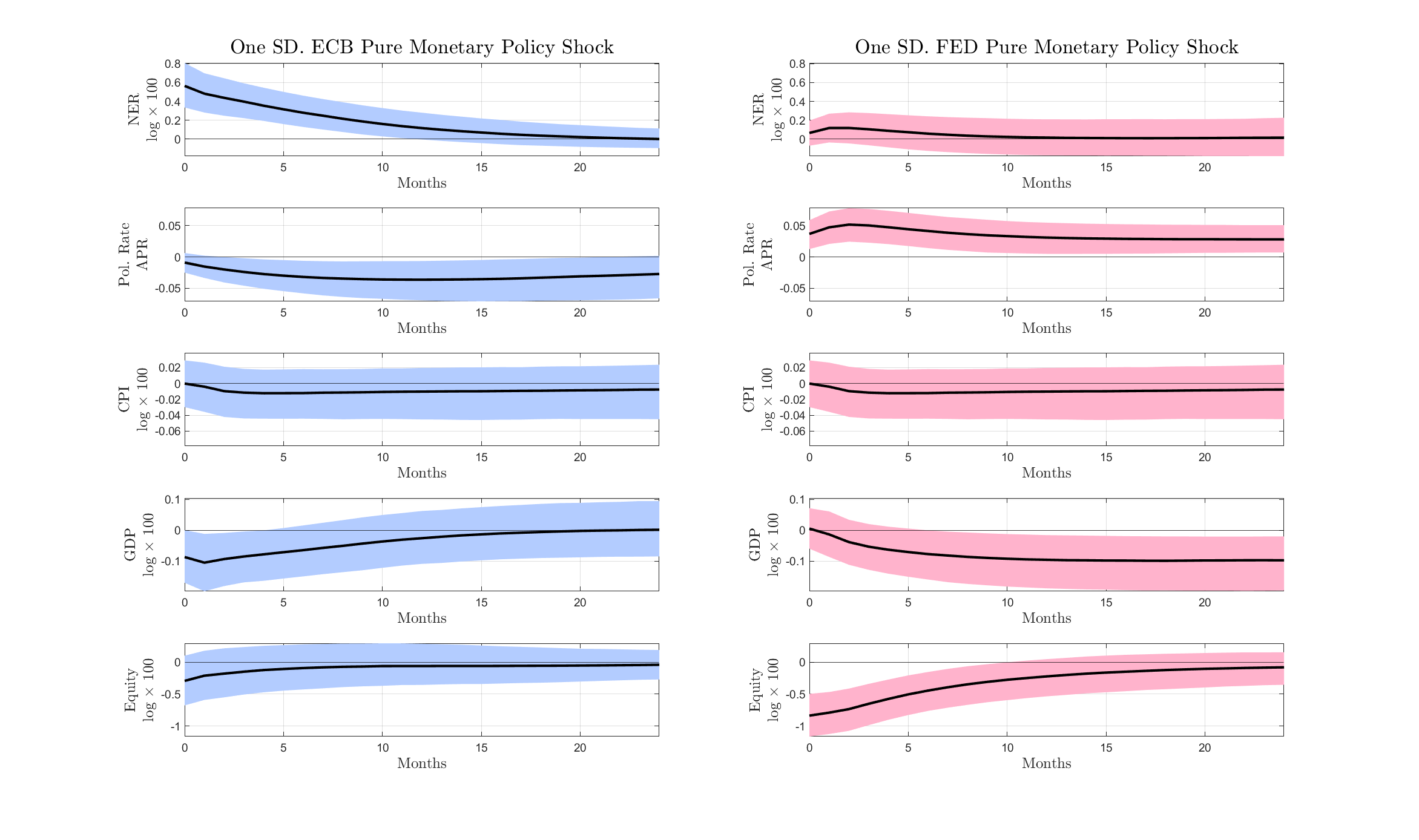}
    \caption{ECB \& Fed Interest Rate Shocks \\ Poor Man’s Sign Restriction}
    \label{fig:Comparison_JK_MP_PM}
    \floatfoot{\textbf{Note:} Left panels correspond to ECB shocks; right panels to Fed shocks. The black line denotes the median IRF; shaded areas represent 90\% posterior credible intervals.}
\end{figure}

\medskip
\noindent
\textbf{Robustness to alternative identification: Poor Man’s Sign Restriction.} 
As a robustness check, we re-estimate the IRFs using the ``Poor Man’s Sign Restriction'' (PMSR) of \citet{jarocinski2020central}, which classifies policy announcements solely by the sign of the co-movement between interest rate and equity surprises: negative co-movements are interpreted as monetary policy shocks, while positive co-movements are classified as information shocks. Figure \ref{fig:Comparison_JK_MP_PM} reports the resulting responses.

The responses under PMSR are qualitatively and quantitatively similar to those obtained under the median-rotation benchmark. ECB shocks continue to generate a depreciation of the Canadian dollar, a decline in GDP, and limited financial spillovers, while Fed shocks remain associated with tighter financial conditions, falling equity prices, and a more gradual but persistent contraction in output. These results confirm that our main findings are robust to alternative identification assumptions.

\begin{figure}[p]
    \centering
    \includegraphics[height=15cm,width=15cm]{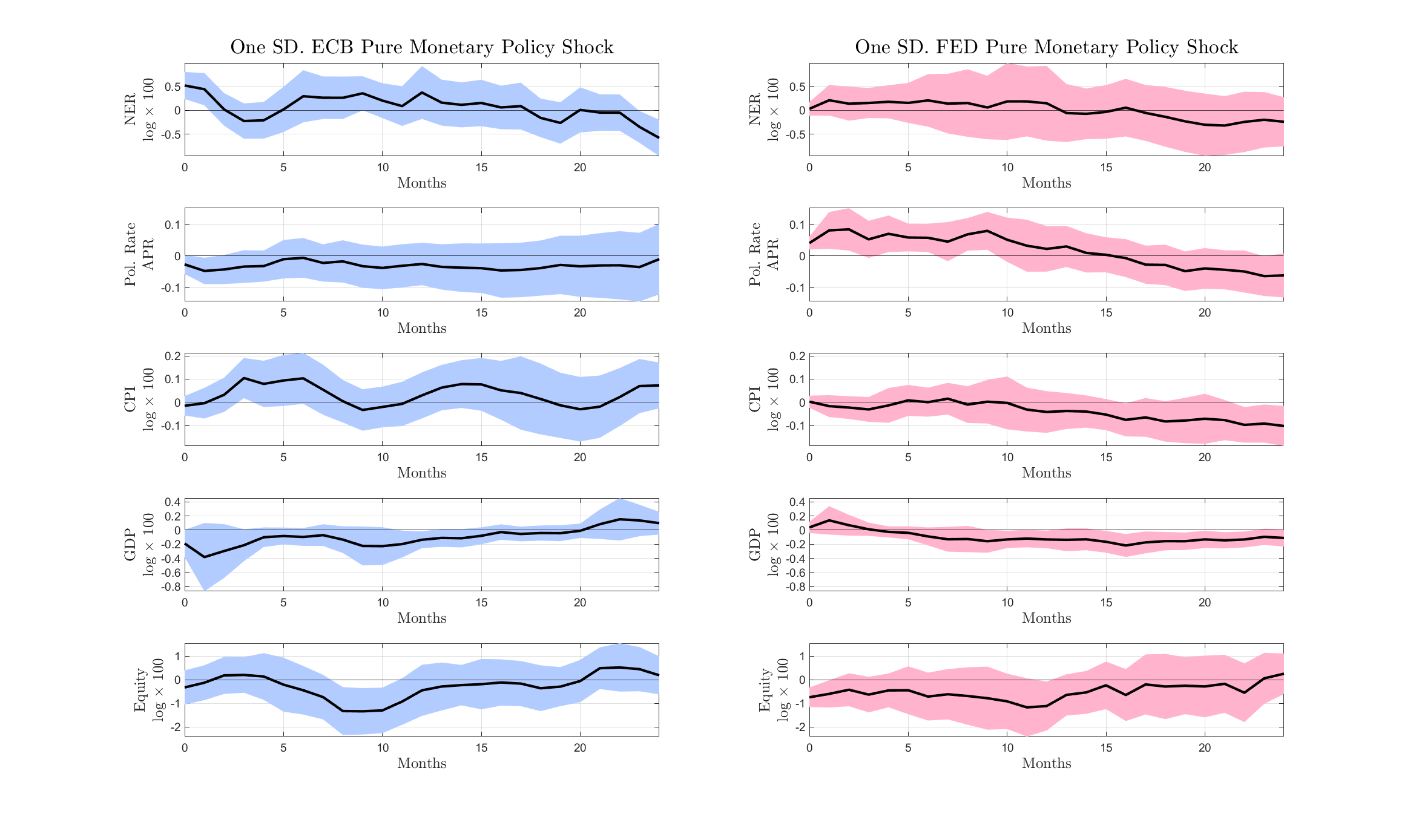}
    \caption{ECB \& Fed Interest Rate Shocks \\ Local Projections — Median Rotational Angle}
    \label{fig:Comparison_JK_MP_Median_LP}
    \floatfoot{\textbf{Note:} Left panels correspond to ECB shocks; right panels to Fed shocks. The black line denotes point estimates from Equation \ref{eq:LP_Benchmark}; shaded areas represent 90\% confidence intervals. Standard errors clustered at the time level.}
\end{figure}

\medskip
\noindent
\textbf{Robustness to estimation method: Local Projections.} 
We also assess robustness to the choice of estimation method. While Bayesian VARs and local projections estimate the same population impulse responses under standard assumptions, they can differ in small samples. Figure \ref{fig:Comparison_JK_MP_Median_LP} reports IRFs estimated using the local projection specification in Equation \ref{eq:LP_Benchmark}.

The results confirm the robustness of our main findings. ECB shocks still generate a depreciation of the Canadian dollar, a fall in the policy rate, and a contraction in GDP. Compared with the VAR estimates, the equity response emerges with a delay, becoming significant only after six months. Fed shocks once again produce tighter financial conditions and a more gradual but persistent decline in GDP, with statistical significance only after a year.

Overall, the local projection estimates validate the robustness of our findings across both identification strategies and estimation methods. They also underscore the central contrast: ECB shocks transmit rapidly through real activity, while Fed shocks operate more strongly through financial conditions with slower real effects. This distinction motivates the more detailed analysis of transmission mechanisms in Section \ref{sec:transatlantic_spillovers_transmission_channels}.

%%%%%%%%%%%%%%%%%%%%%%%%%%%%%%%%%%%%%%%%%%%%%%%%%%%%%%%%%%%%%%%%%%%%%
\section{Differences in Transmission Channels} \label{sec:transatlantic_spillovers_transmission_channels}

This section examines the mechanisms behind the spillovers documented above. Building on the baseline specification, we augment the model with variables that capture Canada’s key trade and financial linkages. Our goal is to test whether the contrasting responses to ECB and Fed shocks reflect different transmission channels: ECB shocks are expected to propagate mainly through real and trade-based channels, particularly the oil sector, which is central to Canadian exports and external vulnerability, whereas Fed shocks are hypothesized to transmit primarily through financial conditions. We first analyze international trade effects, focusing on oil markets, and then turn to financial variables such as interest rates, housing, and credit aggregates.

%%%%%%%%%%%%%%%%%%%%%%%%%%%%%%%%%%%%%%%%%%%%%%%%%%%%%%%%%%%%%%%%%%%%%
\subsection{The Role of Trade \& Oil Prices} \label{subsec:transmission_role_oil}

We begin by examining whether the spillovers documented above operate through Canada’s trade linkages, with a particular focus on the oil sector. Oil is central to the Canadian economy, accounting for roughly 10\% of GDP and 20–25\% of goods exports in recent years. Because oil is globally priced and highly sensitive to international demand, it provides a natural test of trade-based transmission.

\begin{figure}[p]
    \centering
    \includegraphics[height=15cm,width=16cm]{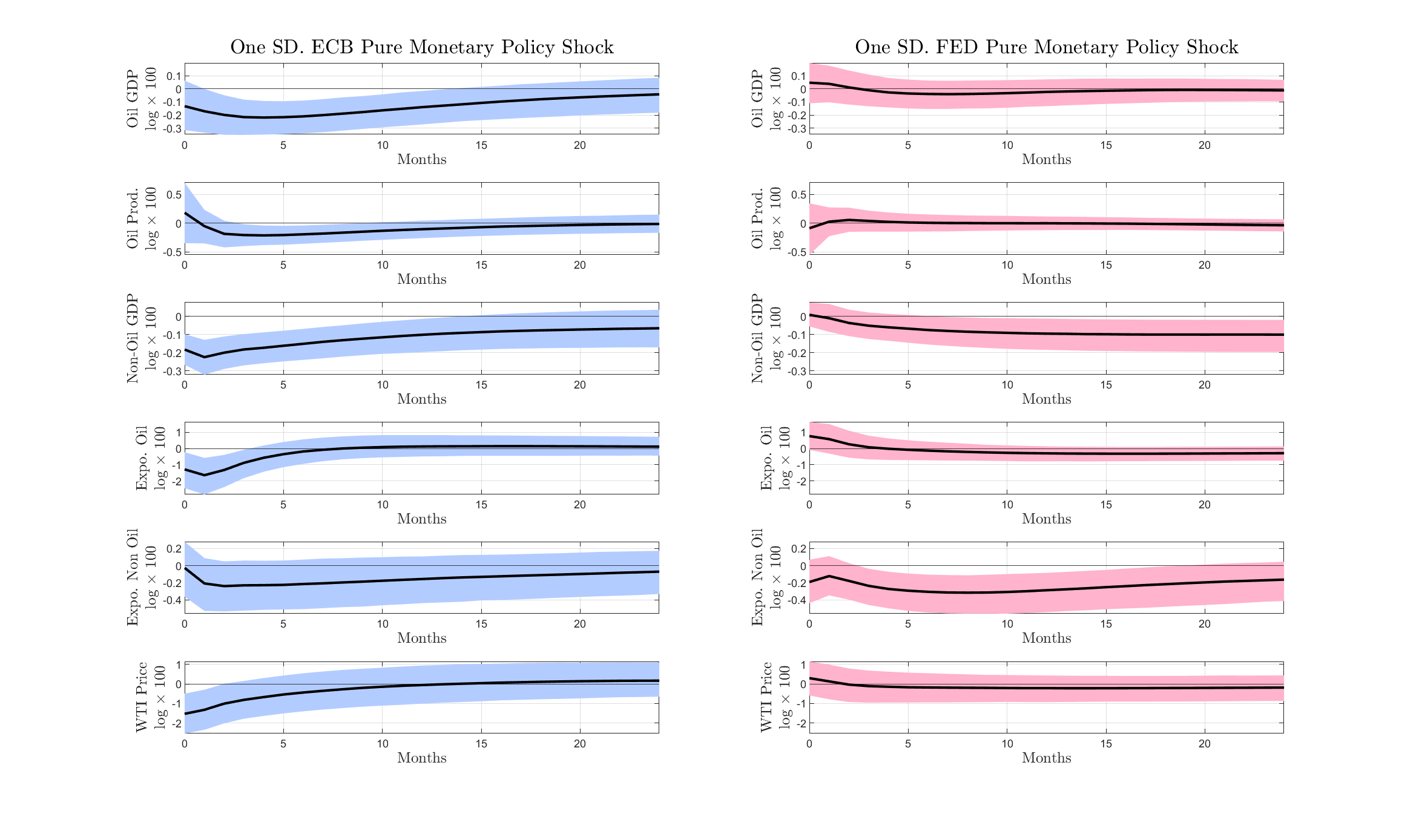}
    \caption{ECB \& Fed Interest Rate Shocks \\ Responses of Oil and Non-Oil Variables}
    \label{fig:Comparison_Oil_vs_Nonoil}
    \floatfoot{\textbf{Note:} Left panels correspond to ECB shocks; right panels to Fed shocks. The black line denotes the median IRF; shaded areas represent 90\% posterior credible intervals.}
\end{figure}

\medskip
\noindent
\textbf{Role of Oil.}  
Figure \ref{fig:Comparison_Oil_vs_Nonoil} reports IRFs from a specification augmented with oil GDP (value added), oil production (in cubic meters), non-oil GDP, oil exports, non-oil exports, and the WTI oil price. 

A contractionary ECB shock leads to a sharp and persistent decline in the WTI price, followed by significant contractions in Canadian oil production, oil value added, and oil exports. Non-oil GDP also falls, while non-oil exports remain broadly unaffected. By contrast, Fed shocks leave oil prices, oil production, and oil exports unchanged, but generate sizable and persistent contractions in non-oil GDP and non-oil exports. These patterns suggest that ECB shocks transmit primarily via global commodity demand, whereas Fed shocks operate through broader aggregate demand and financial conditions.

\begin{figure}[p]
    \centering
    \includegraphics[height=15cm,width=16cm]{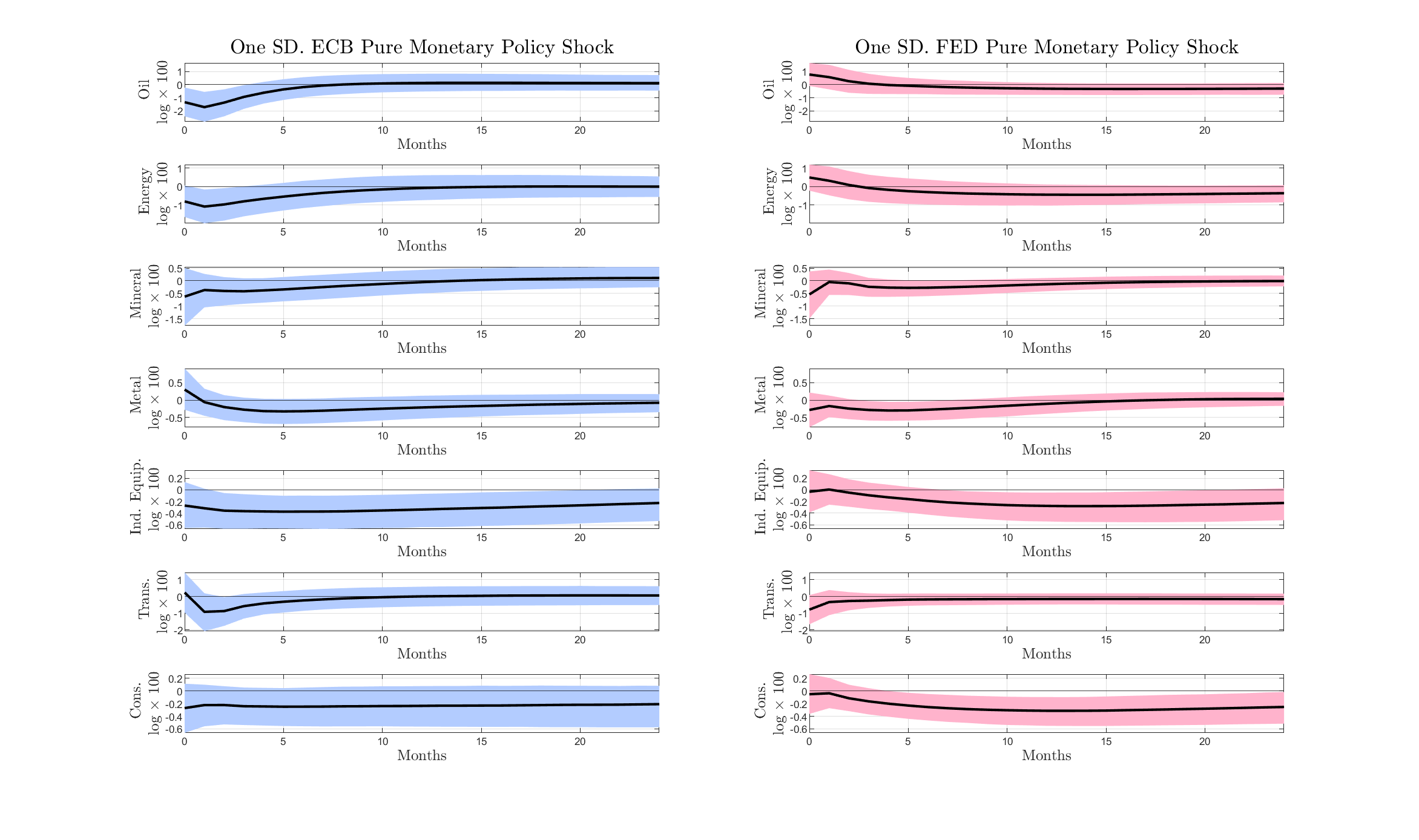}
    \caption{ECB \& Fed Interest Rate Shocks \\ Responses of Canadian Export Categories}
    \label{fig:Comparison_Export_Components}
    \floatfoot{\textbf{Note:} Left panels correspond to ECB shocks; right panels to Fed shocks. The black line denotes the median IRF; shaded areas represent 90\% posterior credible intervals.}
\end{figure}

\medskip
\noindent
\textbf{Export heterogeneity.}  
To probe trade spillovers more finely, we disaggregate exports into seven categories: oil, energy, minerals, metals, industrial equipment, transport goods, and consumption goods.\footnote{Each export component is introduced one at a time into the benchmark specification.} Figure \ref{fig:Comparison_Export_Components} displays the results.

ECB shocks produce sharp declines in energy, metals, and industrial equipment exports, consistent with their impact on global commodity demand. These effects are more gradual than the immediate drop in oil exports, reflecting slower-moving global demand conditions. Transport and consumption goods exports show little response. Fed shocks, in contrast, reduce most non-oil export categories with a lag, consistent with the slower international transmission of tighter U.S.\ monetary policy. Oil and energy exports again show no significant reaction.

\medskip
Taken together, these results underscore significant differences in trade spillovers. ECB shocks generate swift, sector-specific contractions driven by global commodity markets, while Fed shocks propagate more gradually and broadly across non-commodity exports. This highlights the central role of global demand, and especially oil prices, in mediating the spillovers of foreign monetary policy to Canada.

%%%%%%%%%%%%%%%%%%%%%%%%%%%%%%%%%%%%%%%%%%%%%%%%%%%%%%%%%%%%%%%%%%%%%
\subsection{The Role of Financial Variables} \label{subsec:transmission_financial_variables}

While the previous subsection highlighted the importance of trade and commodity channels, particularly for ECB shocks, we now turn to the role of financial conditions. Specifically, we examine how foreign monetary policy shocks affect Canadian interest rates, mortgage rates, and housing prices. This analysis helps assess whether financial markets constitute an independent transmission channel, especially in the case of Fed shocks, where trade effects appear more muted.

\begin{figure}[p]
    \centering
    \includegraphics[height=15cm,width=15cm]{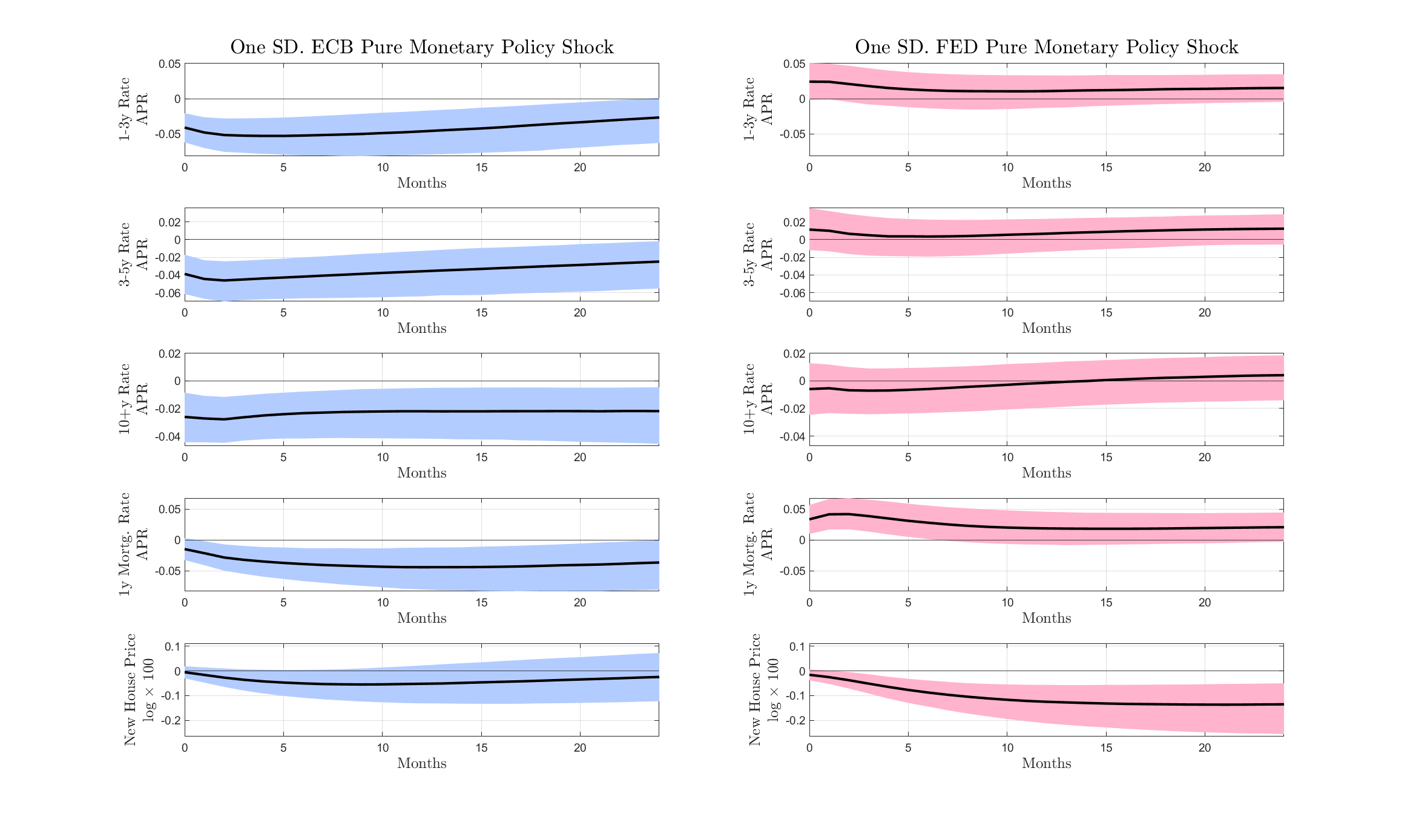}
    \caption{ECB \& Fed Interest Rate Shocks \\ Responses of Canadian Financial Variables}
    \label{fig:Comparison_Financial_Price}
    \floatfoot{\textbf{Note:} Left panels correspond to ECB shocks; right panels to Fed shocks. The black line denotes the median IRF; shaded areas represent 90\% posterior credible intervals.}
\end{figure}

Figure \ref{fig:Comparison_Financial_Price} presents IRFs from a specification augmented with Canadian financial-price variables: government bond yields at 1–3, 3–5, and 10+ year maturities, the 1-year mortgage rate, and the New Housing Price Index. 

ECB shocks are associated with significant declines in Canadian government bond yields across the maturity spectrum, consistent with a dovish response by the Bank of Canada to ECB tightening. At the same time, there is little evidence of international financial spillovers: mortgage rates and housing prices remain essentially unchanged, and longer-term borrowing costs show no signs of upward pressure. In this sense, ECB shocks do not tighten Canadian financial conditions; if anything, they are accompanied by a domestic easing response.

Fed shocks, by contrast, trigger a clear tightening of Canadian financial conditions. Short-term government bond yields rise immediately, reflecting the Bank of Canada’s policy rate increase in response to Fed tightening. Unlike the ECB case, this policy response is reinforced by international financial spillovers: mortgage rates climb significantly, and housing prices decline persistently for more than a year. Medium- and long-term yields also edge up, though less sharply. Overall, Fed shocks spill over into both public and private credit markets, tightening borrowing costs and depressing asset prices in Canada.

Taken together, these results reinforce the distinction between the two sources of spillovers: ECB shocks propagate mainly through trade and commodity channels, whereas Fed shocks transmit through financial conditions, tightening borrowing costs, and depressing asset prices in Canada.

%%%%%%%%%%%%%%%%%%%%%%%%%%%%%%%%%%%%%%%%%%%%%%%%%%%%%%%%%%%%%%%%%%%%%
\subsection{The Impact of ECB Shocks on the U.S. Economy} \label{subsec:transmission_US_economy}

To better understand the channels through which ECB monetary policy affects Canada, we first examine its impact on the U.S.\ economy, Canada’s largest trading and financial partner. If ECB shocks influence U.S.\ output, prices, and financial conditions, they may transmit indirectly to Canada through trade and cross-border linkages.

\begin{figure}[p]
    \centering
    \includegraphics[height=13cm,width=15cm]{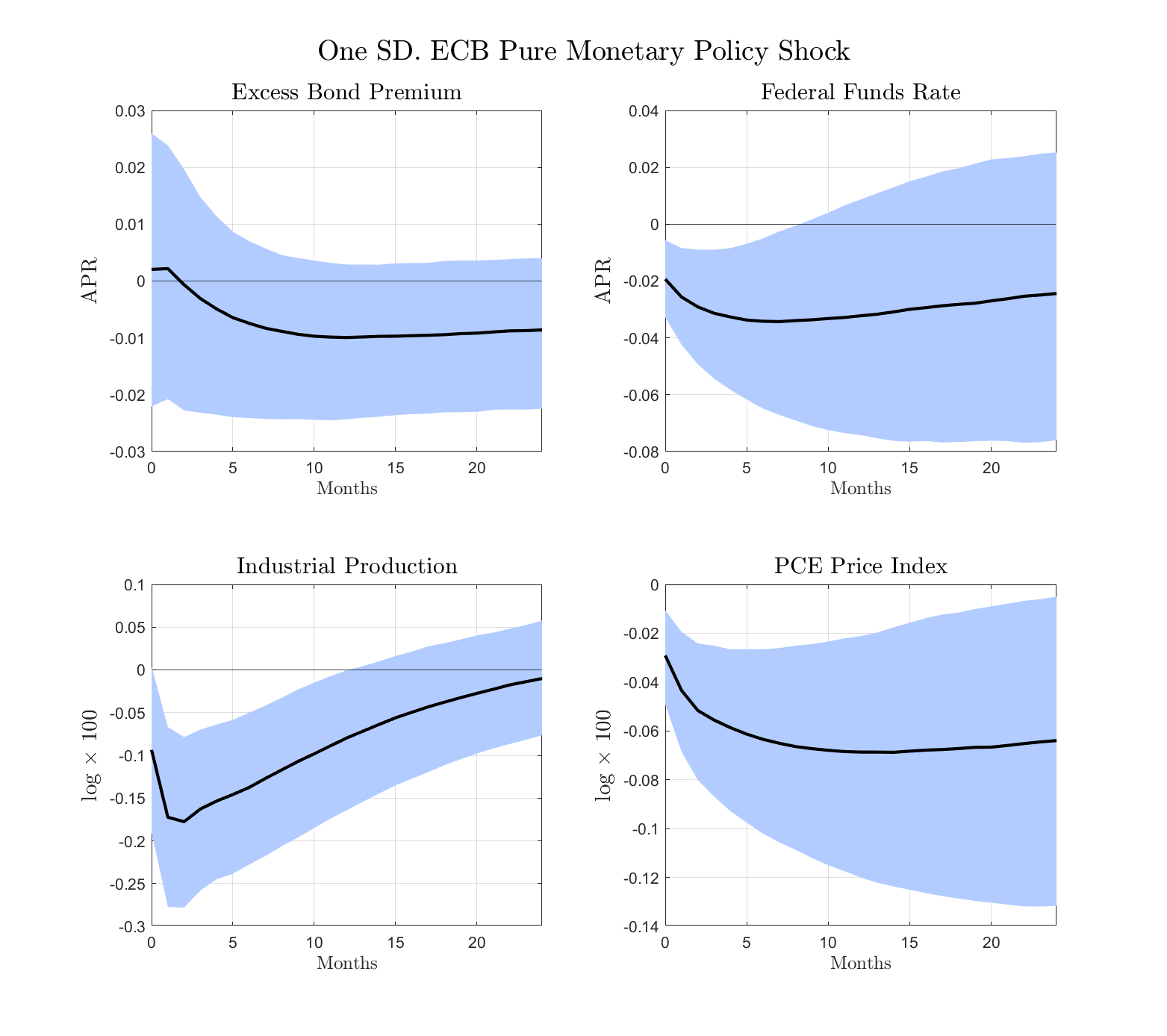}
    \caption{Responses of U.S.\ Macroeconomic Variables to an ECB Shock}
    \label{fig:Comparison_JK_MP_Median_US}
    \floatfoot{\textbf{Note:} One-standard-deviation ECB pure monetary policy shock. Shaded areas denote 90\% posterior credible intervals.}
\end{figure}

Figure \ref{fig:Comparison_JK_MP_Median_US} reports the responses of key U.S.\ macroeconomic and financial variables. A contractionary ECB shock leads to a persistent decline in U.S.\ industrial production and a gradual fall in the PCE price index, consistent with weaker domestic demand. On the financial side, the excess bond premium compresses and the federal funds rate falls, indicating that U.S.\ financial conditions ease in response to the shock. These results suggest that ECB policy has meaningful real and financial spillovers into the U.S.\ economy.

\begin{figure}[p]
    \centering
    \includegraphics[height=13cm,width=15cm]{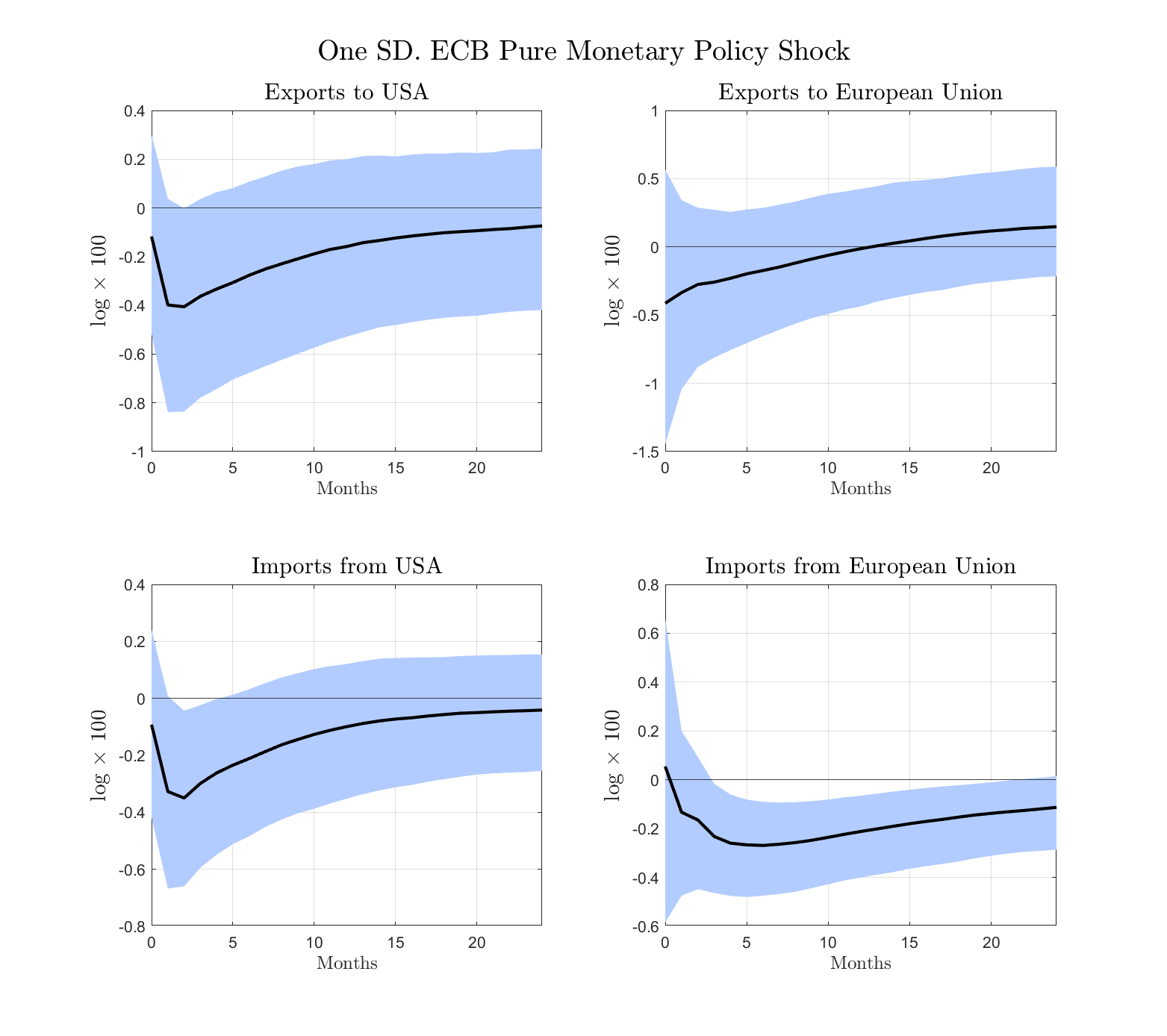}
    \caption{Responses of Canadian Trade Flows to an ECB Shock}
    \label{fig:Comparison_JK_MP_Median_US_Trade}
    \floatfoot{\textbf{Note:} One-standard-deviation ECB pure monetary policy shock. Shaded areas denote 90\% posterior credible intervals.}
\end{figure}

We next examine how these U.S.\ spillovers affect Canadian trade flows. Figure \ref{fig:Comparison_JK_MP_Median_US_Trade} shows that ECB shocks generate significant and persistent contractions in Canadian exports to the United States as well as imports from the United States. By contrast, exports to and imports from the European Union display much smaller and statistically weaker responses. This pattern indicates that the transmission of ECB policy to Canada operates primarily through its effects on U.S.\ demand, rather than directly through bilateral trade with Europe.

Taken together, these results show that ECB shocks spill over into the U.S.\ economy and, through this channel, significantly affect Canadian trade with its largest partner. This indirect route complements the direct trade and commodity mechanisms documented above, underscoring the global reach of ECB monetary policy.

%%%%%%%%%%%%%%%%%%%%%%%%%%%%%%%%%%%%%%%%%%%%%%%%%%%%%%%%%%%%%%%%%%%%%%
\section{Additional Results \& Robustness Checks} \label{sec:additional_robustness_checks}

This section presents a range of additional results and robustness exercises to verify the validity and generality of our main findings in Section~\ref{sec:transatlantic_spillovers}. We first examine heterogeneity across components of Canadian GDP to assess whether the aggregate responses are driven by specific sectors. We then evaluate the role of information effects in monetary policy surprises, highlighting the importance of controlling for information effects around central bank announcements. Next, we consider alternative identification strategies and estimation methods, including variations of the rotation procedure. Finally, we test the robustness of our findings by excluding the COVID-19 period, during which macroeconomic dynamics may have been distorted. 

Across all these exercises, the main transmission patterns remain remarkably stable. ECB shocks continue to propagate primarily through trade and commodity channels, while Fed shocks transmit via financial conditions and tighter credit markets. These additional results confirm that our baseline findings are not sensitive to modeling choices or sample selection, underscoring the robustness of the distinction between ECB- and Fed-driven spillovers into the Canadian economy.

\begin{figure}[ht]
    \centering
    \includegraphics[height=13cm,width=15cm]{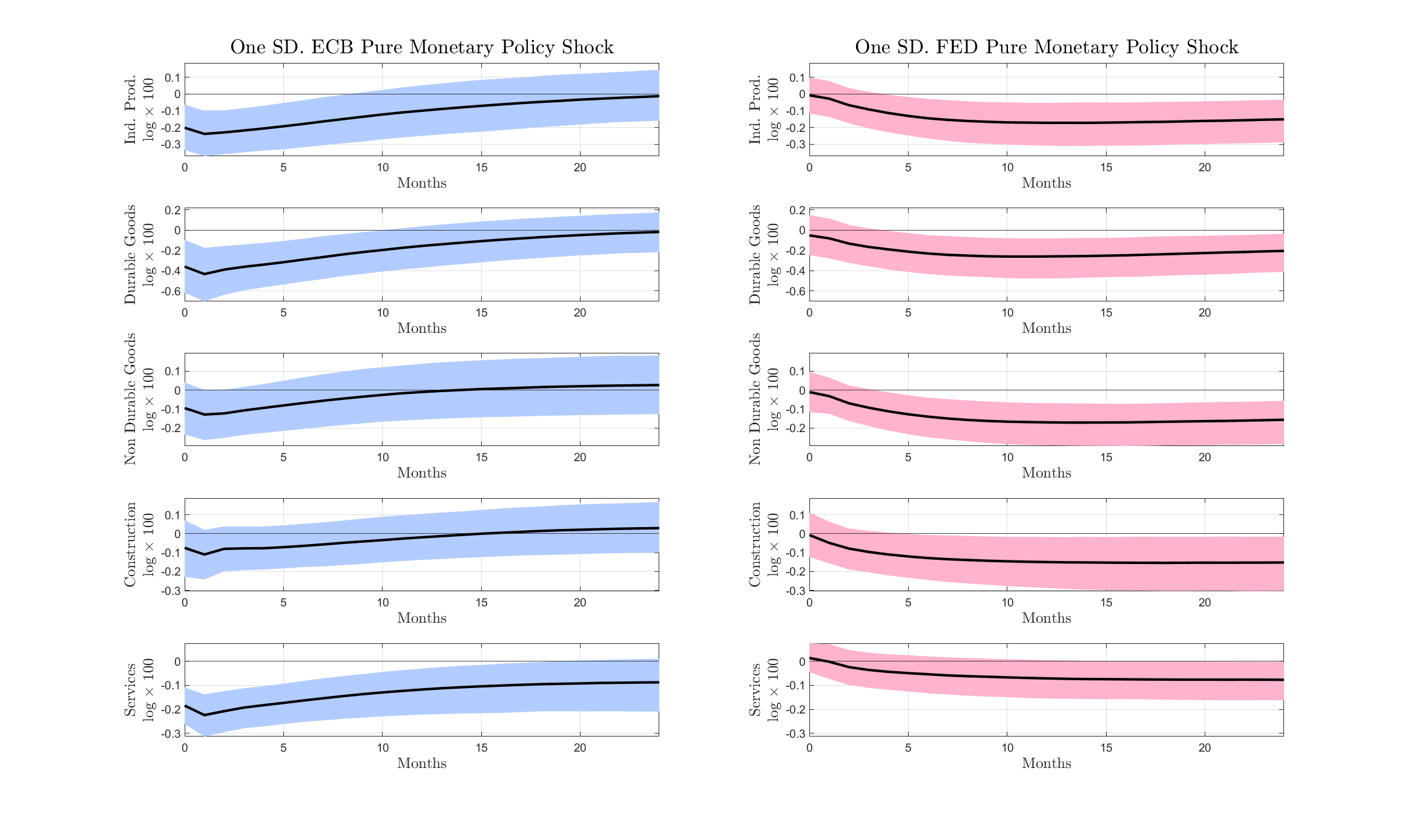}
    \caption{Responses of Canadian GDP Components to ECB and Fed Monetary Policy Shocks}
    \label{fig:Comparison_GDP_Components}
    \floatfoot{\textbf{Note:} Left panels correspond to ECB shocks; right panels to Fed shocks. The black line denotes the median IRF; shaded areas represent 90\% posterior credible intervals.}
\end{figure}

%%%%%%%%%%%%%%%%%%%%%%%%%%%%%%%%%%%%%%%%%%%%%%%%%%
\noindent
\textbf{Impact across GDP components.}  
We next examine whether the responses to foreign monetary policy shocks vary across different components of Canadian GDP. Figure~\ref{fig:Comparison_GDP_Components} reports impulse responses for five key categories: industrial production, durable goods, non-durable goods, construction, and services.

The left column shows the effects of an ECB pure monetary policy shock. The contraction is far from uniform across sectors. Industrial production and durable goods experience a pronounced and persistent decline, larger than in other categories, while services also fall significantly. By contrast, non-durable goods and construction display much weaker responses, with estimates close to zero after only a few months. This heterogeneity is consistent with our earlier results: because ECB shocks do not tighten Canadian financial conditions, credit-sensitive sectors such as construction and consumption-oriented non-durables are relatively insulated, whereas tradable and service industries, more exposed to global demand, bear the brunt of the adjustment.

The right column presents responses to a Fed pure monetary policy shock. In this case, the contraction is more broad-based. Industrial production, durables, and non-durables all decline significantly and persistently, while construction and services exhibit smaller but still noticeable declines. These results align with the financial channel of Fed transmission: tighter borrowing conditions weigh on both goods-producing and service-oriented sectors, resulting in a more widespread slowdown.

Taken together, these findings highlight that the composition of spillovers differs by source. ECB shocks primarily affect tradable goods and services through global demand and commodity channels, while Fed shocks generate broader contractions consistent with financial tightening.

\noindent
\textbf{Biases from Ignoring Information Effects.}  
We next highlight the importance of controlling for the information component of monetary policy announcements. Failing to separate the pure policy action from the information disclosed by central banks risks generating misleading estimates of spillovers. To illustrate this point, we contrast impulse responses to (i) the pure monetary policy shock, (ii) the information shock, and (iii) the standard high-frequency surprise that conflates the two.

\begin{landscape}
\begin{figure}[ht]
    \centering
    \includegraphics[scale=0.55]{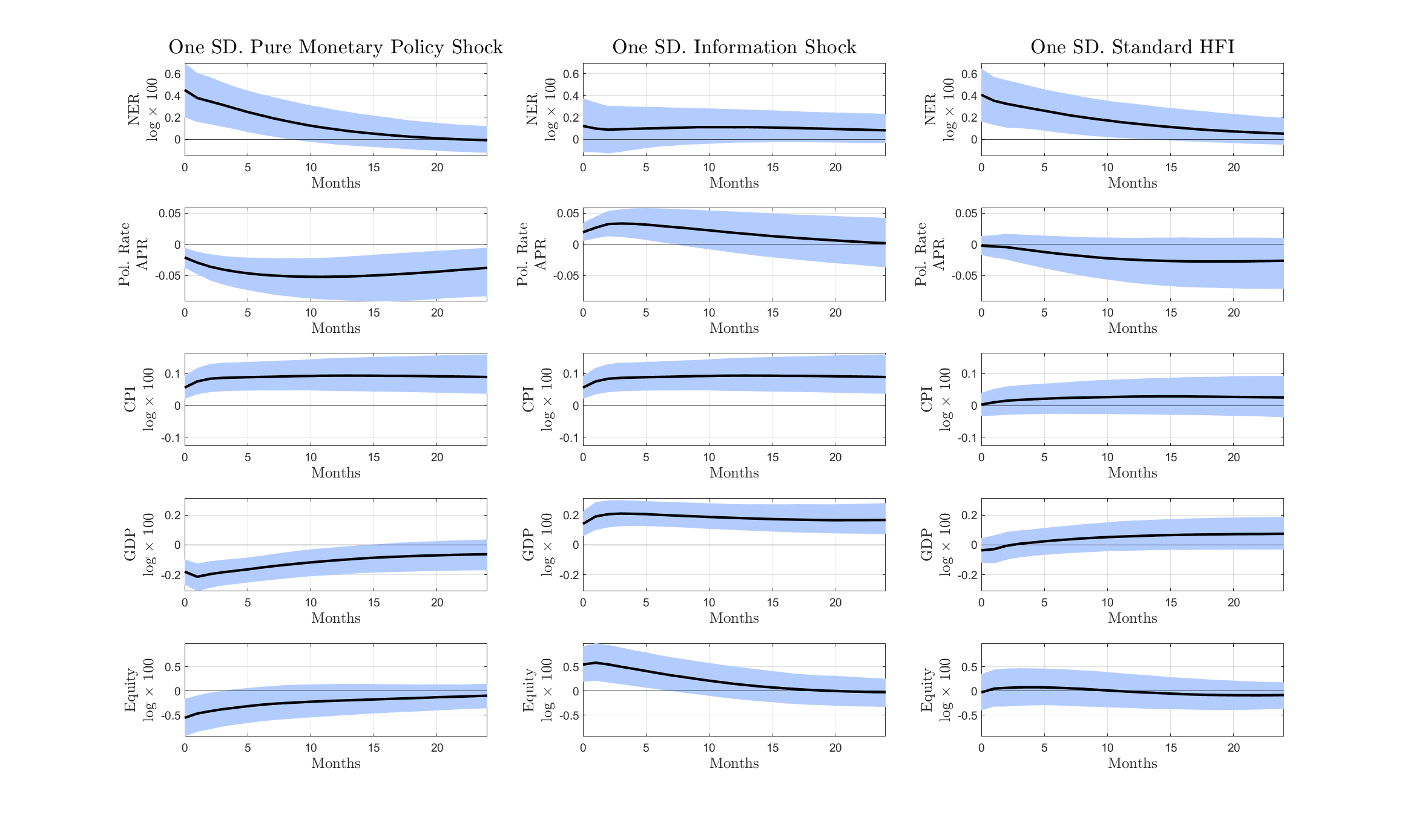}
    \caption{Biases Arising from Information Effects \\ Results for the ECB}
    \label{fig:Bias_ECB_JK_MP_Median}
    \floatfoot{\textbf{Note:} Left column: Pure Monetary Policy Shocks. Middle column: Information Shocks. Right column: Standard HFI surprises. The black line denotes the median IRF; shaded areas represent 90\% posterior credible intervals.}
\end{figure}
\end{landscape}

Figure~\ref{fig:Bias_ECB_JK_MP_Median} presents the results for ECB shocks. The first column shows that a contractionary pure monetary policy shock leads to a depreciation of the Canadian dollar, lower GDP, lower equity prices, and tighter financial conditions. The second column shows that an information shock, when markets revise their expectations of the macroeconomic outlook upward following the announcement, induces the opposite pattern: the exchange rate appreciates, GDP and equity prices rise, and financial conditions ease. 

The rightmost column displays responses to the standard high-frequency identification (HFI) surprise, which combines these two components. As expected, the resulting dynamics fall between those of the pure policy and information shocks. In practice, this means that using the standard HFI surprise without accounting for its informational content systematically biases inference. For example, GDP appears to rise modestly following an ECB rate hike, and equity prices increase briefly on impact, giving the false impression of expansionary spillovers from monetary tightening. 

\begin{landscape}
\begin{figure}[ht]
    \centering
    \includegraphics[scale=0.55]{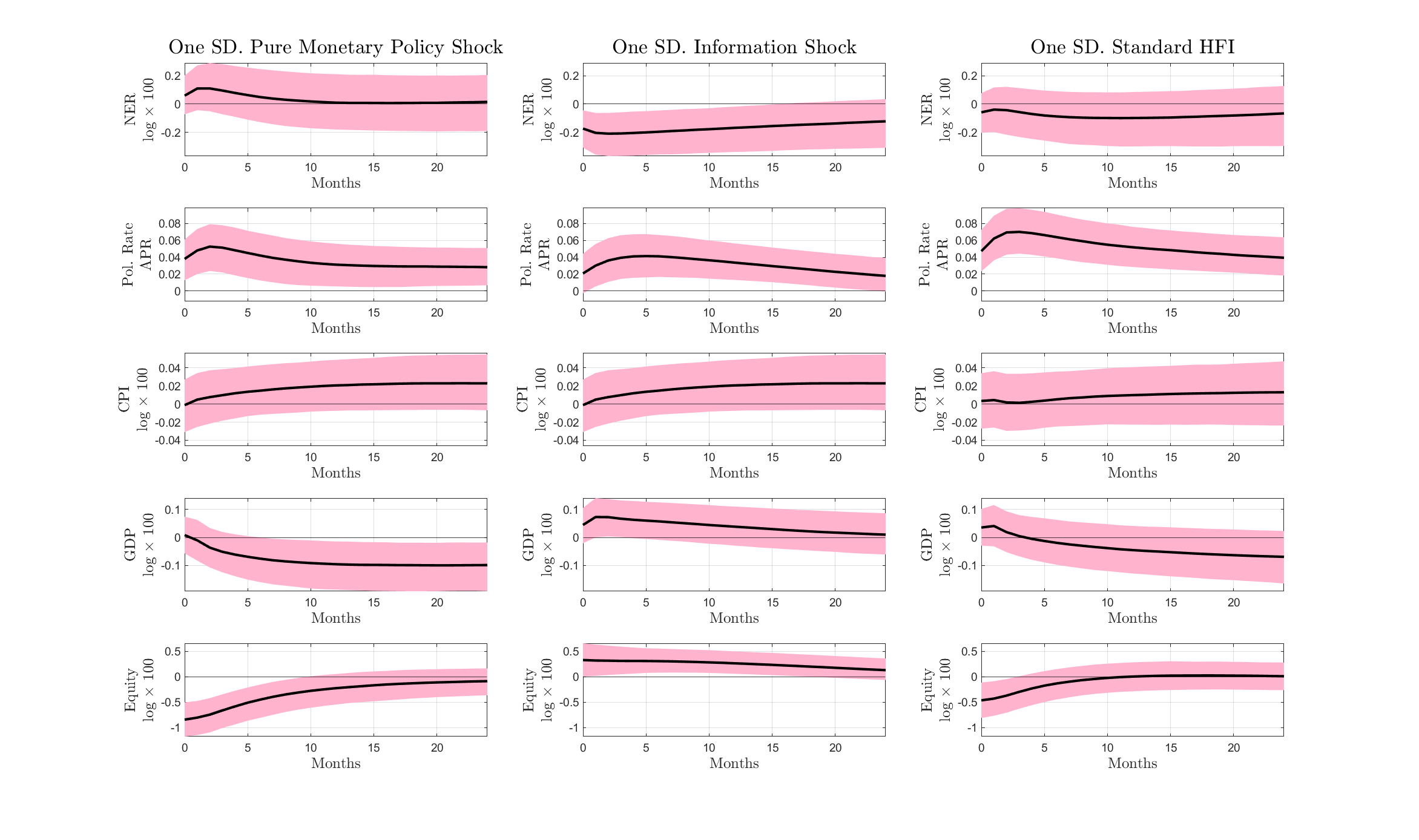}
    \caption{Biases Arising from Information Effects \\ Results for the Fed}
    \label{fig:Bias_Fed_JK_MP_Median}
    \floatfoot{\textbf{Note:} Left column: Pure Monetary Policy Shocks. Middle column: Information Shocks. Right column: Standard HFI surprises. The black line denotes the median IRF; shaded areas represent 90\% posterior credible intervals.}
\end{figure}
\end{landscape}

Figure~\ref{fig:Bias_Fed_JK_MP_Median} reports the corresponding results for the Fed. The same pattern emerges: pure monetary policy shocks generate contractionary spillovers, while information shocks are expansionary, and the conflated HFI surprise lies in between. These results reinforce the point that neglecting information effects leads to biased and potentially misleading conclusions about the international transmission of monetary policy.

\noindent
\textbf{Robustness to the COVID-19 Period.}  
To ensure that our results are not driven by the extraordinary volatility associated with the COVID-19 pandemic, we re-estimate the benchmark specification on a truncated sample that ends in February 2020, prior to the onset of lockdowns and unprecedented policy interventions. This exercise serves two purposes: (i) it avoids the distortions created by an exceptional macroeconomic episode, and (ii) it tests whether our baseline results are robust to the exclusion of this period. 

\begin{figure}[p]
    \centering
    \includegraphics[height=15cm,width=16cm]{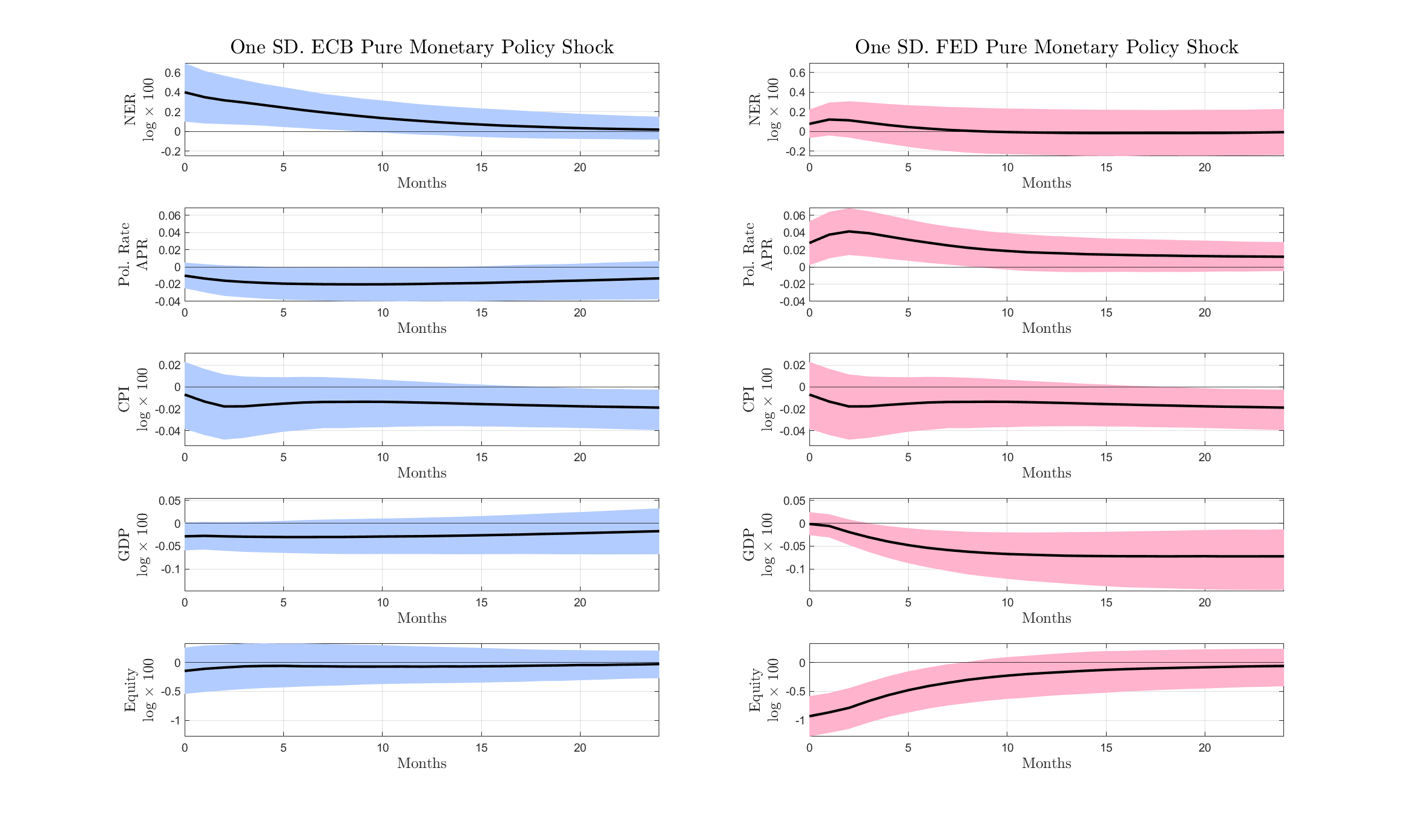}
    \caption{ECB \& Fed Interest Rate Shocks \\ Excluding COVID-19 Period}
    \label{fig:Comparison_JK_MP_Median_Covid}
    \floatfoot{\textbf{Note:} Left panels correspond to ECB shocks; right panels to Fed shocks. The black line denotes the median IRF; shaded areas represent 90\% posterior credible intervals.}
\end{figure}

Figure~\ref{fig:Comparison_JK_MP_Median_Covid} shows that the estimated impulse responses are strikingly similar to those obtained with the full sample. The magnitude, timing, and persistence of responses remain broadly unchanged for both ECB and Fed shocks. This confirms that our main findings are not an artifact of the COVID-19 period and reflect more general features of the international transmission of monetary policy. 

\clearpage
\clearpage

\noindent
\textbf{Robustness to admissible rotations.}  
Our benchmark relies on the median admissible rotation from the \citet{jarocinski2022central} decomposition. To test robustness to this choice, we draw uniformly from the entire set of admissible rotations that satisfy the sign restrictions. For each draw, we recover the associated shock series, re-estimate the VAR, and compute impulse responses. Repeating this procedure many times allows us to construct credibility bands that combine posterior uncertainty with identification uncertainty from the rotation.\footnote{Following \citet{jarocinski2022central}, we normalize the set of admissible rotations to a fixed number of angles (999 in our case), and then take repeated random draws from a uniform distribution over this set.}

\begin{figure}[p]
    \centering
    \includegraphics[height=15cm,width=16cm]{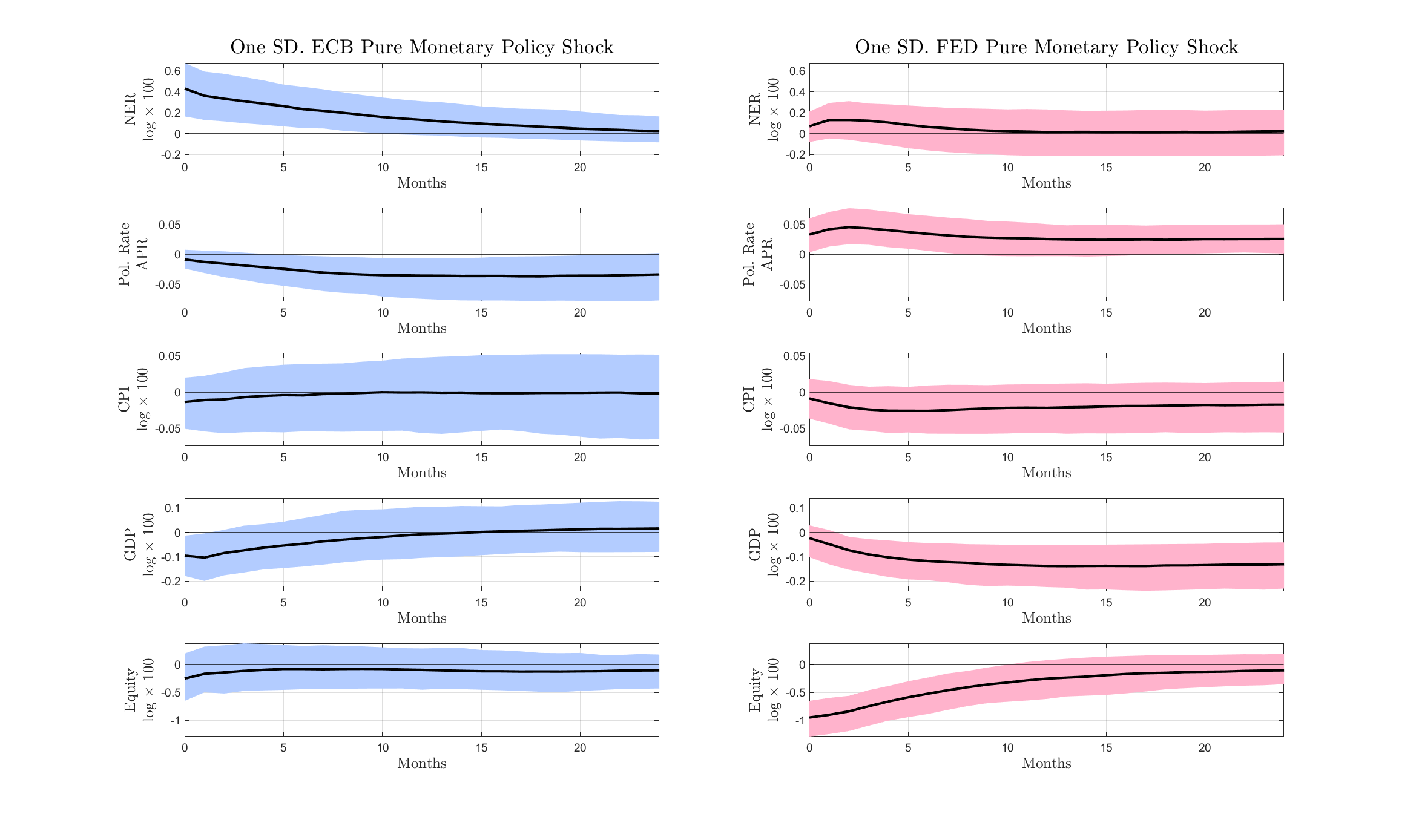}
    \caption{Impulse Responses Integrating Over Admissible Rotations}
    \label{fig:Comparison_JK_MP_Uncertainty_Bounds}
    \floatfoot{\textbf{Note:} Bands are pointwise 90\% credible intervals that account for estimation uncertainty and identification uncertainty from drawing uniformly over all admissible rotational angles. The solid line is the median-rotation benchmark.}
\end{figure}

Figure~\ref{fig:Comparison_JK_MP_Uncertainty_Bounds} shows that the resulting impulse responses closely resemble our benchmark estimates. ECB shocks continue to depreciate the Canadian dollar, are followed by policy easing in Canada, and generate a contraction in activity. Fed shocks remain associated with tighter financial conditions, lower equity prices, and a persistent fall in GDP. While credibility intervals widen slightly when accounting for identification uncertainty, the benchmark responses lie comfortably within the 90\% envelopes. This confirms that our findings are robust to the choice of admissible rotation. 

\medskip
\noindent
\textbf{Summary.}  
Across all robustness exercises, whether examining heterogeneity across GDP components, controlling for information effects, varying identification strategies, altering estimation methods, or excluding the COVID-19 period, the core transmission patterns remain remarkably stable. ECB shocks consistently transmit through trade and commodity markets, while Fed shocks propagate through financial conditions. These results underscore that our main conclusions are not sensitive to modeling choices or sample selection, providing strong evidence for the asymmetric nature of foreign monetary spillovers into Canada.

%%%%%%%%%%%%%%%%%%%%%%%%%%%%%%%%%%%%%%%%%%%%%%%%%%%%%%%%%%%%%%%%%%%%%%
\section{Conclusion} \label{sec:conclusion}

This paper provides new evidence on the international spillovers of foreign monetary policy by documenting how European Central Bank (ECB) interest rate shocks affect the Canadian economy. Using high-frequency data and a sign-restriction-based identification strategy that separates pure monetary policy shocks from information effects, we estimate the responses of Canadian macroeconomic and financial variables to both ECB and Federal Reserve (Fed) shocks, allowing for a systematic comparison of transmission channels.

The results challenge the conventional U.S.-centric view of Canada’s external dependence. We show that ECB shocks generate sizable spillovers into Canada, despite limited direct trade and financial linkages with the euro area. These spillovers operate primarily through real channels—most importantly global oil prices and international trade, whereas Fed shocks transmit predominantly through financial conditions, tightening domestic interest rates, raising borrowing costs, and depressing equity and housing markets. This asymmetry reveals that Canada is simultaneously exposed to distinct foreign monetary transmission mechanisms: global demand shocks from Europe and financial shocks from the United States.

A contractionary ECB monetary tightening reduces global oil prices, leading to contractions in Canadian oil production and exports, as well as a decline in GDP. The Canadian dollar depreciates temporarily, while the Bank of Canada eases domestic policy in response. By contrast, Fed tightening triggers persistent financial tightening, with broad declines in credit-sensitive sectors and more gradual effects on trade. These findings underscore the importance of recognizing heterogeneity in the transmission of foreign shocks.

We demonstrate that these results are robust to alternative identification strategies, different econometric specifications, and the exclusion of the COVID-19 period. They also hold when examining subcomponents of GDP and trade, and they remain valid when accounting for uncertainty across admissible rotations. Crucially, we show that failing to control for the informational content of monetary policy announcements biases inference, even in a small open economy like Canada.

Our findings carry broader implications for both research and policy. For scholars, they highlight the need to move beyond a U.S.-only view of foreign monetary spillovers and to recognize the role of multiple global central banks. For policymakers, they emphasize that small, commodity-exporting economies remain vulnerable to international shocks through both trade and financial channels, and that these vulnerabilities depend on which central bank is the source of disturbance. Future research should investigate whether similar asymmetries characterize the exposure of other small open economies and how domestic monetary frameworks can best incorporate the global nature of monetary transmission.

For Canadian policymakers in particular, these results underscore the challenge of conducting monetary policy in an open economy: even when domestic financial linkages with the euro area are limited, external shocks from the ECB can spill over forcefully through global demand and commodity markets, while Fed shocks operate through financial conditions.

%%%%%%%%%%%%%%%%%%%%%%%%%%%%%%%%%%%%%%%%%%%%%%%%%%%%%%%%%%%%%%%%%%%%%%%%%%%%%%%%%%%%%%
\newpage
\bibliography{main.bib}

\newpage
\appendix

%%%%%%%%%%%%%%%%%%%%%%%%%%%%%%%%%%%%%%%%%%%%%%%%%%%%%%%%%%%%%%%%%%%%%%%%%%%%%%%%%%%%%%
%%%%%%%%%%%%%%%%%%%%%%%%%%%%%%%%%%%%%%%%%%%%%%%%%%%%%%%%%%%%%%%%%%%%%%%%%%%%%%%%%%%%%%

\end{document}